\documentclass[sn-mathphys]{sn-jnl}

\usepackage{amssymb}
\usepackage{graphicx}
\usepackage{epstopdf}
\usepackage{amsmath}
\usepackage{epsfig}
\usepackage{color}
\usepackage{ulem}
\usepackage{stmaryrd}
\usepackage{amsmath}

\jyear{2021}

\theoremstyle{thmstyleone}

\theoremstyle{thmstyletwo}

\theoremstyle{thmstylethree}

\begin{document}

\title[Using "AI Poincaré" to analyze non-linear integrable optics]{Using "AI Poincaré" to analyze non-linear integrable optics}

\author*[1]{\fnm{Lazare} \sur{Osmanov}}\email{lazare.osmanov1521@gmail.com}
\author*[2]{\fnm{Nilanjan} \sur{Banerjee}}

\affil*[1]{\orgdiv{School of Physics}, \orgname{Free University of Tbilisi}, \orgaddress{\street{David Aghmashenebeli Alley}, \city{Tbilisi}, \postcode{0159}, \country{Georgia}}}

\affil*[2]{\orgdiv{IOTA}, \orgname{Fermilab}, \orgaddress{\street{N Kirk Rd}, \city{Batavia}, \postcode{60539}, \country{USA}}}

\abstract{This study dives into the applicability of using automated discovery of conserved quantities in dynamical systems relevant to accelerator physics. Specifically, we explore the performance of AI Poincaré in analyzing numerical trajectory data obtained using the McMillan system of non-linear integrable optics. A comprehensive evaluation of the algorithm's performance is conducted through diverse methodologies. These include the analysis of the estimated number of conserved quantities embedded in a dataset and the deviation of interpolated points on the inferred manifold with respect to points in actually in the dataset. the investigation identifies an optimal range of perturbation distances where the underlying manifold extraction algorithm inside AI Poincaré exhibits optimal performance. Additionally, an improved neural network architecture is proposed based on the observed results. Finally, we apply the algorithm to preliminary experimental data from the Integrable Optics Test Accelerator at Fermilab to successfully infer the number of conserved quantities even in the presence of fast decoherence of the measured signal.}


\maketitle

\section{Introduction}
The first step in the design of all existing particle accelerators has been to consider linear optics only. In this approximation, the beam dynamics is described by linear differential equations \cite{c4}.

\begin{equation}
\label{a12}
x'' + \Big(K(s) + \frac{1}{\rho_x^2(s)}\Big)x \&= \frac{\delta}{\rho_x(s)}\\
\end{equation}

\begin{equation}
\label{a12}
y'' - K(s)y \&= 0
\end{equation}

where $x$ and $y$ are the horizontal and vertical positions, respectively, of a single particle with respect to a reference trajectory of the beam as it travels through the vacuum system. $s$ and $\delta$ represent the longitudinal position and the deviation of the longitudinal momentum relative to the design value, respectively. $\rho_x (s)$ and $K(s)$ are the radius of curvature of the trajectory due to the dipoles and the focusing due to the quadrupoles, respectively.

\begin{figure}[t]
\centering\includegraphics[width=8cm]{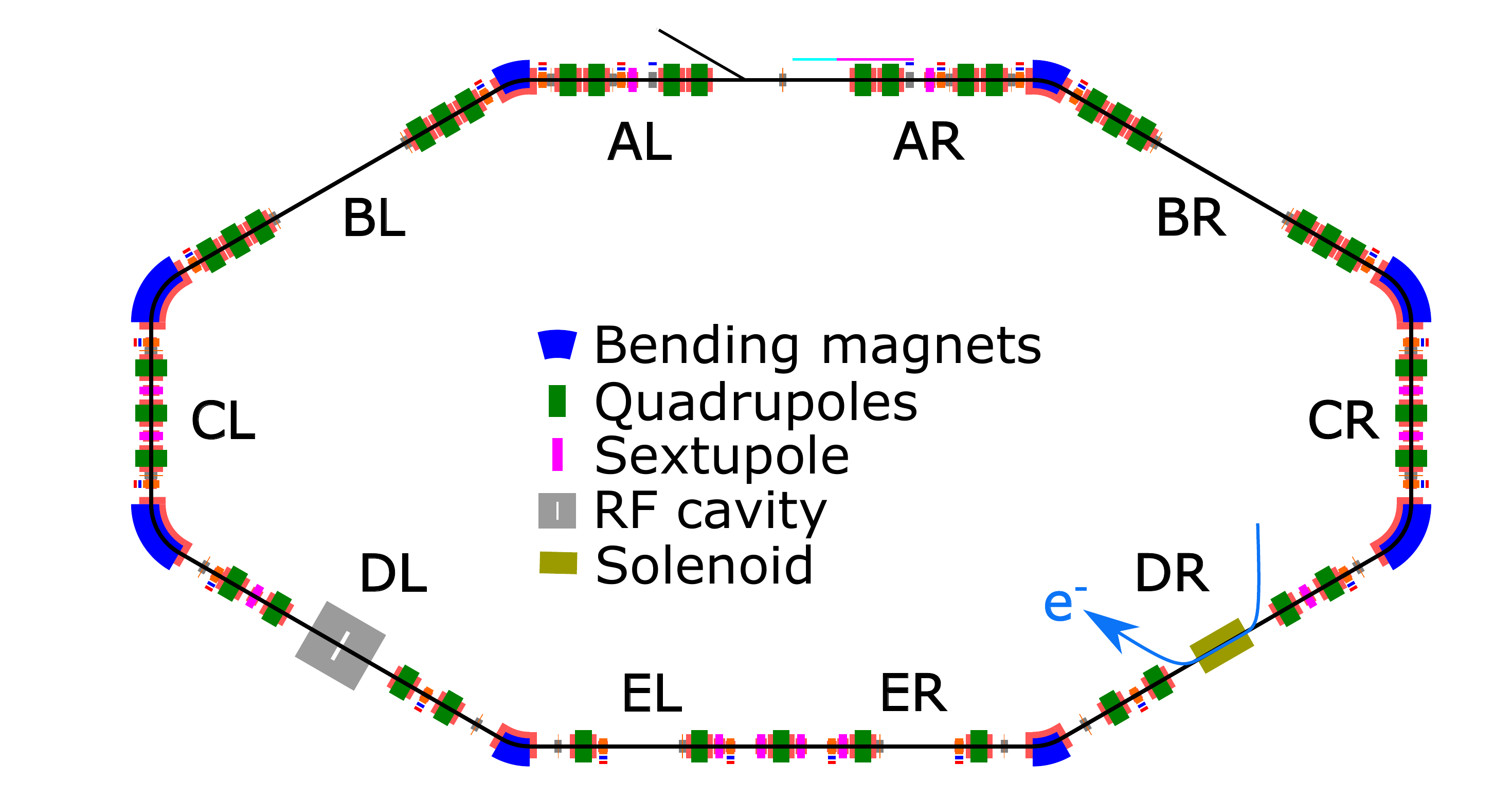}
\caption{Schematics of FermiLab integrable optics acceleration "IOTA" \cite{c4}}\label{fig1}
\end{figure}

\begin{figure}[t]
\centering\includegraphics[width=8cm]{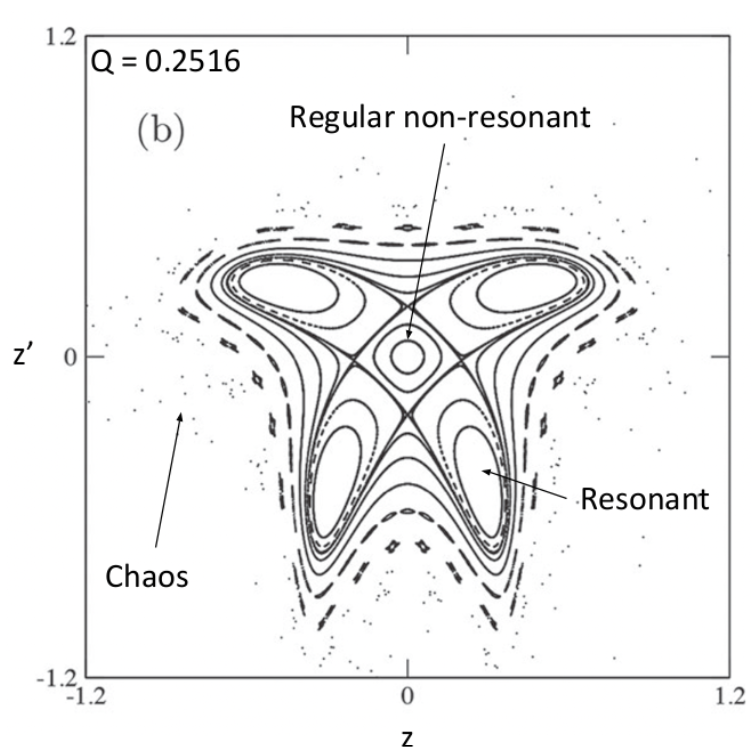}
\caption{Graphical representation of the IOTA (Integrable Optics Test Accelerator) ring showcasing its structural layout. The Danilov-Nagaitsev non-linear element is placed in section BR, while the McMillan kick will be realized as an electron lens incorporated in section DR.}\label{fig2}
\end{figure}

In a perfect world, we wish that every accelerator could be described in a linear way As an example, a schematic diagram of the Integrable Optics Test Accelerator at Fermilab is shown in Fig.\ref{fig1} . The bending magnets denoted in blue generate $\rho_x$, while the quadrupoles shown in green generate $K$ in Eq. (1). 
However all real accelerators are subject to non-linear dynamics because of tolerances in the fabrication of real magnets, and the inclusion of dedicated non-linear elements for compensation of chromaticity and damping of collective instabilities. Unfortunately, non-linearity often leads to chaos and chaos tends to manifest in large amplitude excursions of the particles with respect to the center of the beam pipe, and this leads to beam loss. Fig.\ref{fig2} shows a projection of a single-particle trajectory in phase-space for a toy model of an accelerator with a second order non-linearity with different initial conditions. This illustrates three major classes of dynamics observed in accelerator beams: regular non-resonant, resonant and chaotic motion. We design particle accelerators to operate in a regime where most particles remain in the non-resonant part of phase space. However in practice, this limits the maximum number of particles or beam intensity which the machine supports. Hence a method to increase the volume contained in the regular non-resonant region of phase space or eliminate the chaotic region entirely would be hugely beneficial to increasing the beam intensity supported by future accelerators.

Non-linearity doesn't immediately mean chaotic dynamics since non-linear systems still could be integrable The Integrable Optics Test Accelerator (IOTA) at Fermilab is a storage ring dedicated to beam physics research on non-linear dynamics, cooling and many other topics. The flagship experiment at this facility is the demonstration of non-linear integrable optics (NIO) using the Danilov-Nagaitsev system \cite{c5} which introduces two conserved quantities in the transverse motion of single particles.

Unlike chaotic systems which can be bounded but never closed, the phase-space of integrable systems, will always be bounded and closed as well. FIn accelerators, NIO results in a phase-space where single particles travel on non-intersecting manifolds which are topologically equivalent to concentric tori filling phase-space., so even if the particle is perturbed randomly at some point of its path in the lattice, it will go to another, infinitesimally different tori. Topologically, the tori are the same and hence the distance between the original and the perturbed trajectory will not diverge over time, thus avoiding chaos. Only a few known non-linear integrable optics systems exist for use in particle accelerators: McMillan \cite{c6}, Danilov-Nagaitsev and the octupole string system, but with the development of new automatic manifold learning algorithms such as "Neural Empirical Bayes" \cite{NEB} we can explore the possibility of automated searches for new approximate NIO systems for particle accelerators.

In the next sections, we introduce AI Poincaré\cite{AIP} and demonstrate its operation using the McMillan system. Then, we analyze the performance of the manifold learning algorithm and optimize the neural network structure, also we test how AI Poincare works for real experimental data. Finally, we summarize our findings and outline the next steps.

\section{Description of AI Poincare}
AI Poincare is a machine learning algorithm that estimates the number of invariants conserved in numerical trajectory data of a dynamical system. If the number of conserved quantities is the same as the degrees of freedom of the system then the dynamics is integrable, thus increasing the volume of phase-space where the dynamics is regular non-resonant.

AI Poincare consists of three steps, the first step is \textit{pre-whitening} which refers to scaling the data so that the distribution of points is isotropic in phase-space and this also removes the trivial, linear conserved quantities in the system. Linear invariants are removed from the data using Principal Component Analysis cite{c7} This process involves calculating the orthogonal basis vectors representing the data and then measuring the relative information content or \textit{explained ratio} corresponding to each of these eigen-directions. The directions with the least explained ratio represent linear invariants and are discarded. 
The second step is \textit{Monte-Carlo sampling}, where random points $\mathbf{x}_i$ are chosen from the data set, perturbed with random noise $\mathbf{n}_i$, of zero mean and rms length scale $L$ and then we construct a neural network $P_\theta$ to pull the perturbed point $\mathbf{y}_i ={x}_i +  \mathbf{n}_i$ back to ${x}_i$.
As the training progresses, the neural network learns the global shape of the manifold and is thus able to pull back points $\mathbf{y}_i$ back to the manifold. We use the RMS deviation as a definition of loss function $Loss = \sqrt{\frac{1}{n} \sum_{i=1}^{n} x_i^2}$ and "Adam optimizer" \cite{adam} for dynamic "learning rate" parameter optimization. After the training session, We save all the parameters (weights and biases) of Neural Networks.

Then the algorithm chooses a subset of points on the manifold from the original phase-space dataset. Then it applies multiple random perturbations to each point, with a rms length scale matching the walk length scale which was used for the training step of the neural network. Finally, algorithm uses the trained neural network to pull back the points to the manifold. The pulled back points will form a local tangent plane in the neighborhood of the starting point
We can now infer the local dimensionality of the manifold from the total number of linear invariants conserved within this set of points. Hence we apply PCA to the points constituting the local tangent plane and use the same condition $\lambda ... < \epsilon$ to identify the number of local invariants 

We applied AI Poincare to the 2D McMillan map which describes the turn-by-turn transverse phase-space coordinates of a single particle circulating in an accelerator ring, in the presence of a special non-linear element. Knowing that this map preserves exactly two independent invariants of motion, we can study what AI Poincaré is doing under the hood. (appendix)

\begin{figure}[t]
\centering\includegraphics[width=8cm]{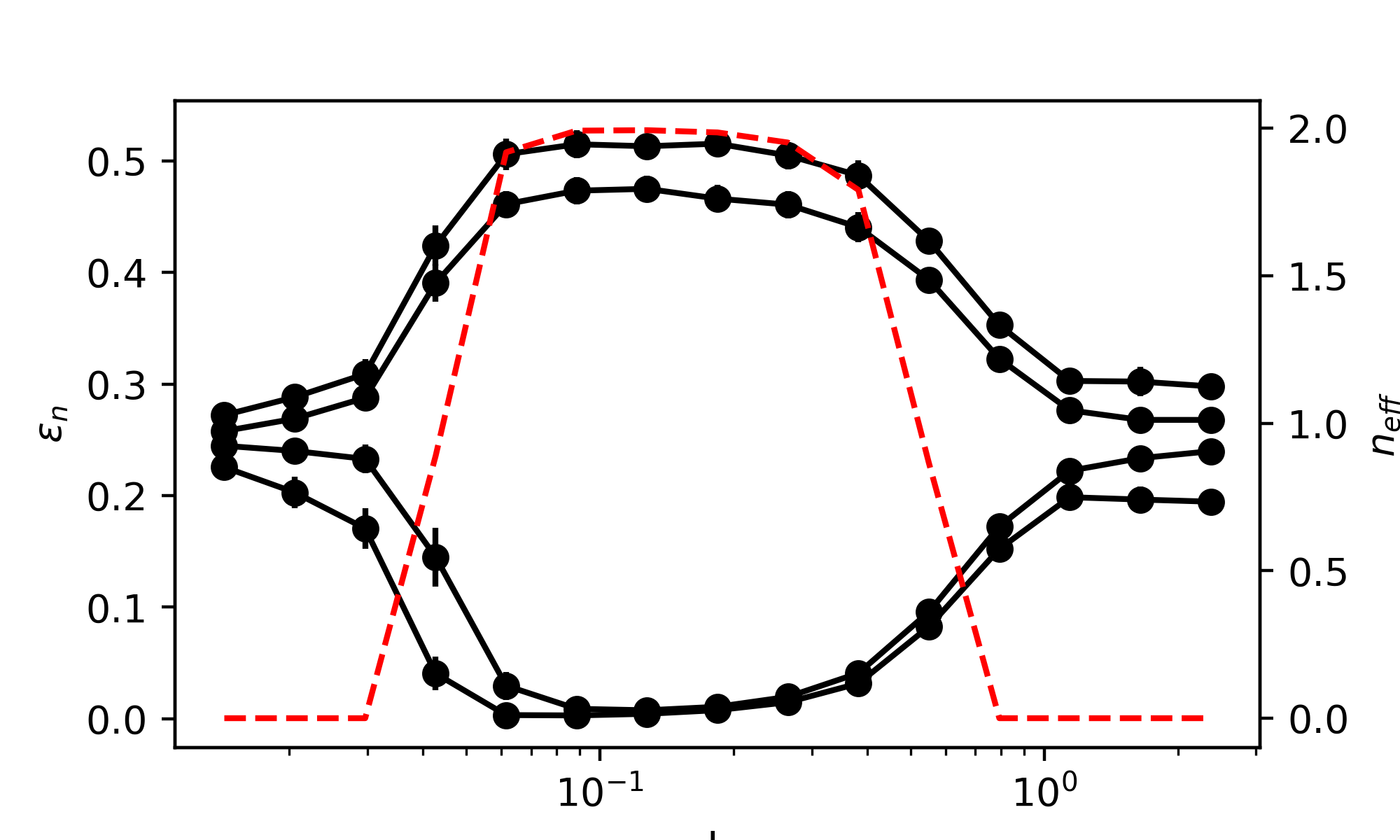}
\caption{ Explained ratio diagram obtained by applying AI Poincaré to a typical turn-by-turn phase-space dataset generated numerically for McMillan lens system}\label{exp}
\end{figure}

Figure~\ref{exp} shows the \textit{explained ratio diagram} for a dataset generated using the McMillan map. Each black line in the plot shows the explained ratio of one phase-space dimension as a function of the random walk length scale during the training process. A small value of explained ratio can be interpreted as a detection of a conservation law. When the neural network is trained using very small walk length scales, it tends to learn the location of the specific points in the data set and the estimated dimensionality is simply $1/N$, where $N$ is the number of phase-space dimensions. At length scales larger than the typical separation between points in the dataset, the training process fails to teach the neural network the correct points to pull back to, and hence the inferred dimensionality is undefined. However, at intermediate walk length scales, the neural network learns the global shape of the manifold. Hence, the dimensionality of the local tangent planes reconstructed using the neural network is globally consistent. This is clearly seen for $L \in [0.1, 0.3]$ in Fig.~\ref{exp}. Since two of the four dimensions have suppressed explained ratios, this implies that there are two invariants in the system, which is consistent with our expectations.

\section{Validating the fitted manifold}
In the field of accelerator physics, AI Poincaré can have two distinct uses: to optimize accelerator optics in order to increase the phase-space volume within which the particle dynamics can be regular non-resonant and to analyze experimental data from non-linear integrable optics system, such as in IOTA. We evaluate the suitability of AI Poincaré to these use cases by looking at (1) how accurately can it learn the structure of the manifold and (2) whether we can sample the global shape of the manifold from the trained neural network. We explore these requirements using the known McMillan system. We can benchmark the accuracy of the Neural Empirical Bayes network by evaluating the invariants on the points pulled back to the manifold. The distance of the pulled-back points from the points in the phase space dataset indicates whether the algorithm learned unknown parts of the manifold not sampled in the dataset.

\begin{figure}[t]
\centering\includegraphics[width=8cm]{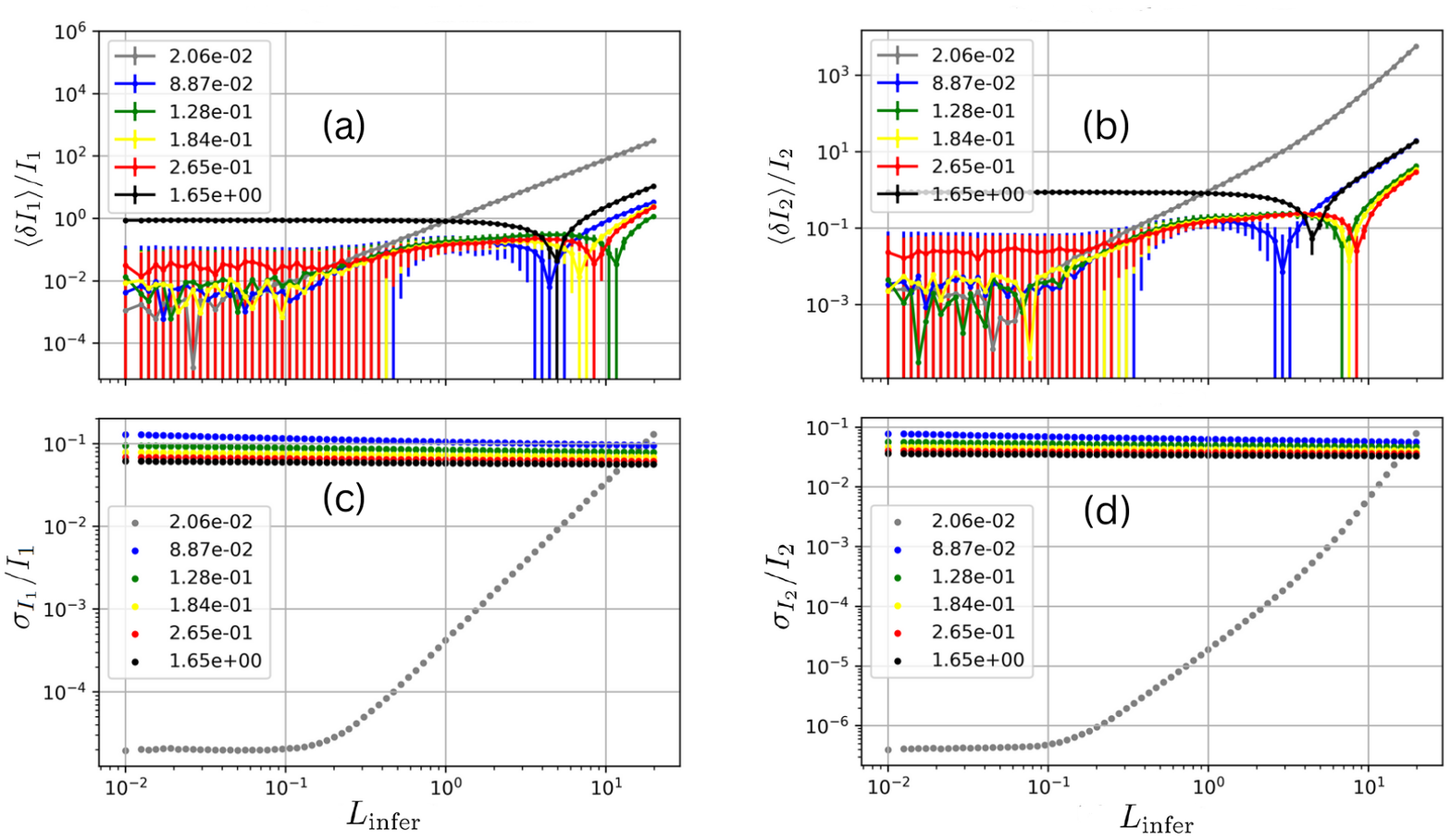}
\caption{Relative deviation of invariants computed from points $\mathbf{x}' = P_\theta{\mathbf{y}}$ inferred from the pull network. Panels (a) and (b) depict relative error of the invariants $I_1$ and $I_2$ with respect to the original values, respectively, as a function of the rms walk length scale used during inference. Panels (c) and (d) show the uncertainty of the relative error predictions for invariants  $I_1$ and $I_2$ respectively.}\label{fig4}
\end{figure}

The explained ratio diagram displayed in Fig.~\ref{exp} shows the dimensionality of the local tangent planes inferred in the neighborhood of randomly chosen points on the manifold. The normalized length $L$ refers to both the random walk length scale used to train the pull network and also the length scale of the local tangent plane used to estimate the dimensionality of the manifold. To validate the global structure of the manifold learned by $P_\theta$, we compute the change in the value of the invariants of the inferred points compared to that of the original points, i.e. $\delta I_1/I_1 = I_1(P_\theta(\mathbf{x}'))/I_1(\mathbf{x}) - 1$, where the rms distance between the perturbed point $\mathbf{x}'$ and the original point $\mathbf{x}$ on the manifold is $L_\mathrm{infer}$. Better preservation of invariants over large perturbation length scales will indicate that the network indeed encodes the \textit{global} shape of the manifold.

\begin{figure}[t]
\centering\includegraphics[width=8cm]{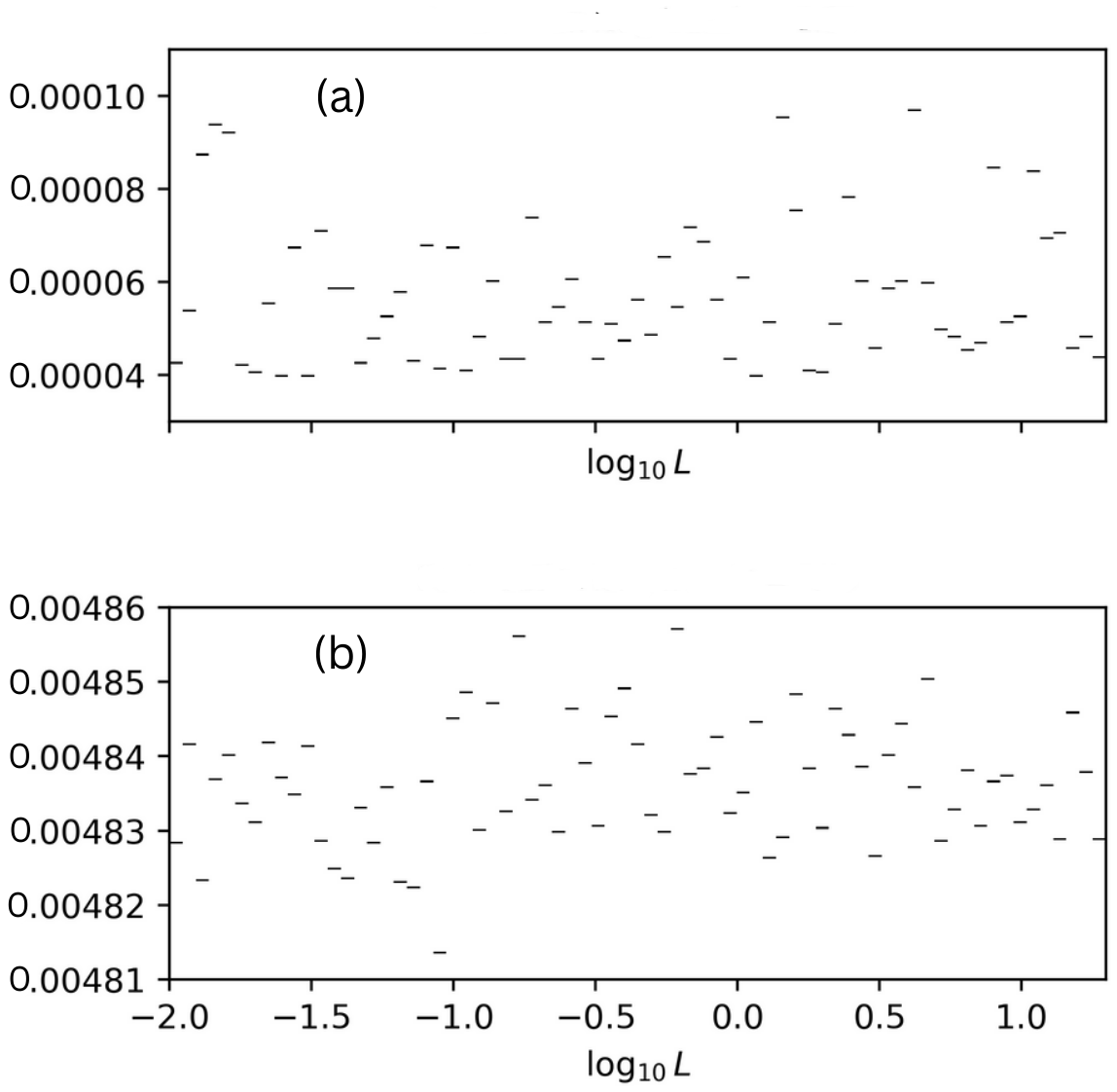}
\caption{ Representation of the displacement between the original point and corresponding pulled-back points for two distinct Neural Networks. The networks were individually trained for different perturbation lengths, on panel (a) the perturbation length $L_1=1.65$ and on panel (b) $L_2=0.13$. }\label{fig3}
\end{figure}

The relative change in both invariants of the McMillan system for points inferred by pull networks, trained at different length scales is shown in Fig.~\ref{fig4}. The gray lines in all panels correspond to the network trained at the smallest normalized length scale of $L=2.06\times10^{-2}$. For $L_\mathrm{infer} \lesssim 0.2$, this network is able to pull back the perturbed points very close to their original location with very low uncertainty, but the performance very quickly degrades for larger length scales. This is consistent with our previous hypothesis, that for small values of $L$, the network learns the location of individual points rather than a global shape.  For the networks trained at intermediate length scales, denoted by the colors, blue, green, yellow, and red, the relative deviation of the invariant values stay below 0.1 for $L_\mathrm{infer} <  1$. This verifies that the network is able to pull back perturbed points onto the manifold even if the points are further away, indicating knowledge of the global shape. The network trained with a length scale greater than the rms size of the manifold, completely fails to pull back to the manifold as evidenced from the relative error curves (black) in panels (a) and (b) which are of order unity.

An additional intriguing aspect of our analysis is examining how far the neural network pulls points from their originally selected locations. This is particularly insightful because a large displacement would indicate that the network has a strong global understanding of the manifold. To investigate this, we constructed a histogram Fig.\ref{fig3} depicting the distribution of pulled-back points for two different perturbation length scales:$L_1=0.13$, $L_2=1.65$. The results demonstrate that, for both perturbation length scales, the displacements remain small, meaning that the neural network predominantly pulls points back to locations very close to their original positions in phase space. This suggests that the network does not fully capture the global structure of the manifold.

\section{Testing AI Poincare on real experimental data}
Our analysis of AI Poincaré using synthetic data from the McMillan map verifies the validity of the manifold fitting and detection of invariant conservation. However, it lacks non-ideal features typical in experimental data. Experiments typically measure the turn-by-turn motion of the centroid of a particle bunch traversing a ring, after it is perturbed from a steady state. The collective motion of a bunch suffers from decoherence, which leads to a decay in the measured signal as a function of time. As an example, panel (a) of Fig.~\ref{dneff} shows turn-by-turn centroid position data measured in IOTA during an experiment with the Danilov-Nagaitsev magnet. Clearly, the amplitude of the position signal decays to noise after $\sim 200$~turns. This implies that quantities which are invariant when considering single particle dynamics decay in time when measuring centroid position of a bunch. Indeed, plugging in the phase-space coordinates into the analytical expressions which should be invariant, shows exponential decay as seen in panel (b) of Fig.~\ref{dneff}. Fortunately, after the manifold is encoded in the pull network, the dimensionality estimation in AI Poincaré happens locally and so the result should be unaffected by decoherence. We test this hypothesis by applying the algorithm on the transverse turn-by-turn phase-space data obtained in IOTA.

\begin{figure}[t]
\centering\includegraphics[width=8cm]{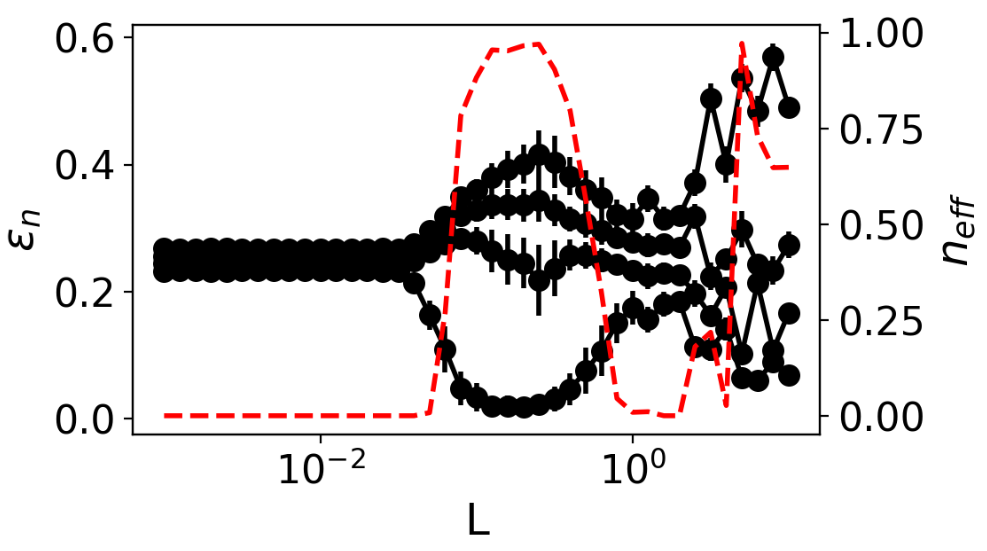}
\caption{ Explained ratio diagram obtained by applying AI Poincaré to a typical turn-by-turn phase-space dataset measured over 200~turns using the Danilov-Nagaitsev magnet in IOTA. The analysis implies that there is only one invariant in the data.}\label{fig5}
\end{figure}

\begin{figure}[t]
\centering\includegraphics[width=8cm]{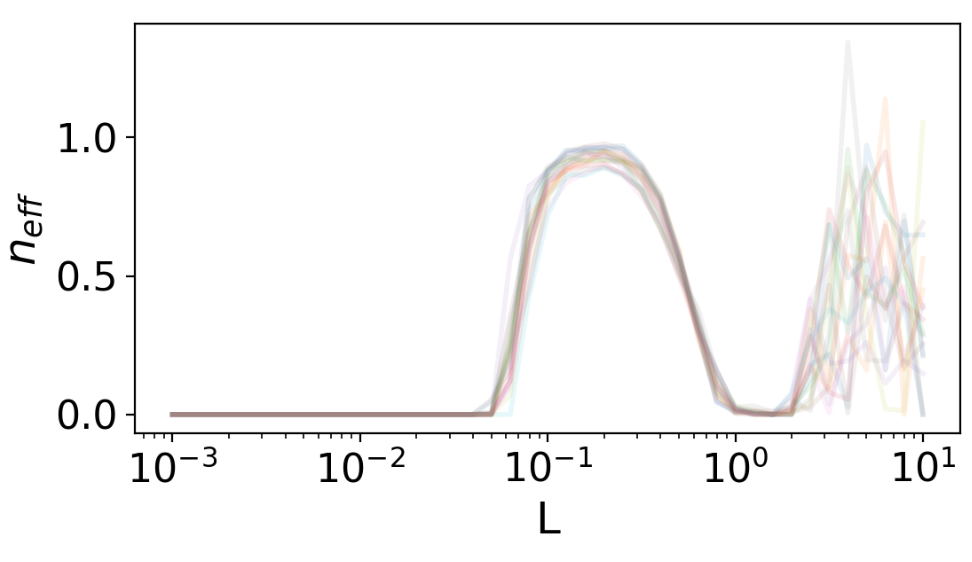}
\centering\includegraphics[width=8cm]{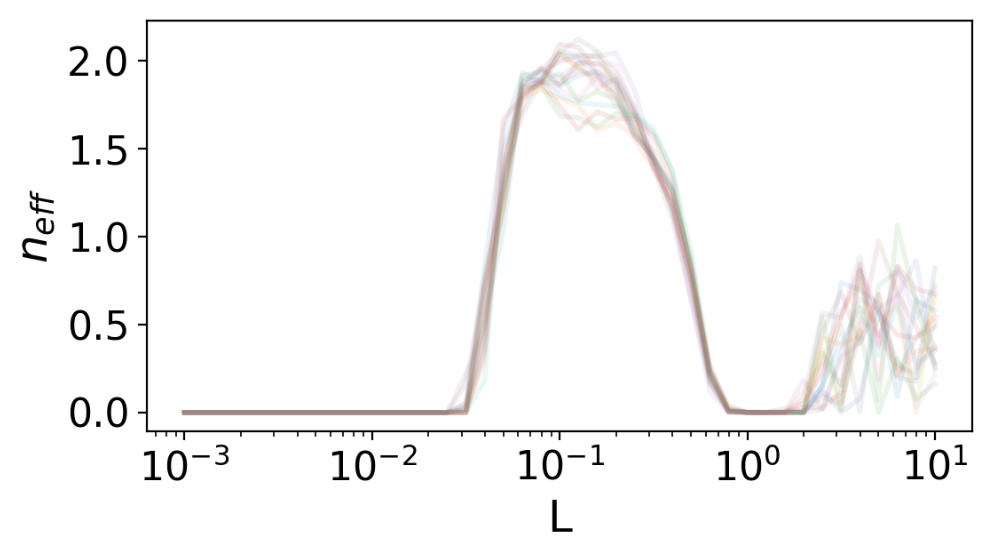}
\caption{ The top graph illustrates the dimensionality of the manifold and the count of conserved quantities over 200 turns of the particle, denoted by $N_{\text{eff}}=1$. In the bottom graph, a distinct perspective is presented, showcasing the manifold dimensionality and the count of conserved quantities for the initial 50 turns of the particle. This comparative analysis reveals a nuanced evolution, suggesting the emergence of an additional conserved quantity within the first 50 turns, a phenomenon not evident over the entire 200-turn trajectory.}\label{dneff}
\end{figure}

The results of applying AI Poincaré to a typical measurement of turn-by-turn position data is shown in Fig.~\ref{fig5}. The plot of effective number of invariants (red) as a function of random walk length scale indicates the conservation of a single invariant. We extend this analysis, to the complete dataset containing 18 separate measurements with different starting perturbations, but the same strength of the non-linear magnet. The effective number of invariants in all datasets as seen in panel (a) of Fig.~\ref{dneff} is 1. This is consistent with the observation that only one of the two invariants plotted in panel (b) of Fig.~\ref{dneff} shows a relatively smooth exponential decay, while the other is very noisy indicating that the function value is not constant even over short timescales. Repeating the same analysis on a subset of the data, containing only the first 50~turns, results in panel (b) of Fig.~\ref{dneff}, which indicates that the number of conserved quantities is 2, which is the theoretical expectation. While more analysis is required to verify this claim, if true, this clearly demonstrates the utility of using AI Poincaré in analyzing integrability in real experimental data.

\section{Optimum Network design for NEB}
We endeavored to optimize the neural network by systematically reducing the associated loss functions. This optimization pursuit was undertaken with the dual objective of enhancing result fidelity and expanding the permissible range of perturbation distances wherein the algorithm performs effectively. We endeavored to optimize the neural network by systematically reducing the associated loss functions. This optimization pursuit was undertaken with the dual objective of enhancing result fidelity and expanding the permissible range of perturbation distances wherein the algorithm performs effectively.

\begin{figure}[t]
\centering\includegraphics[width=8cm]{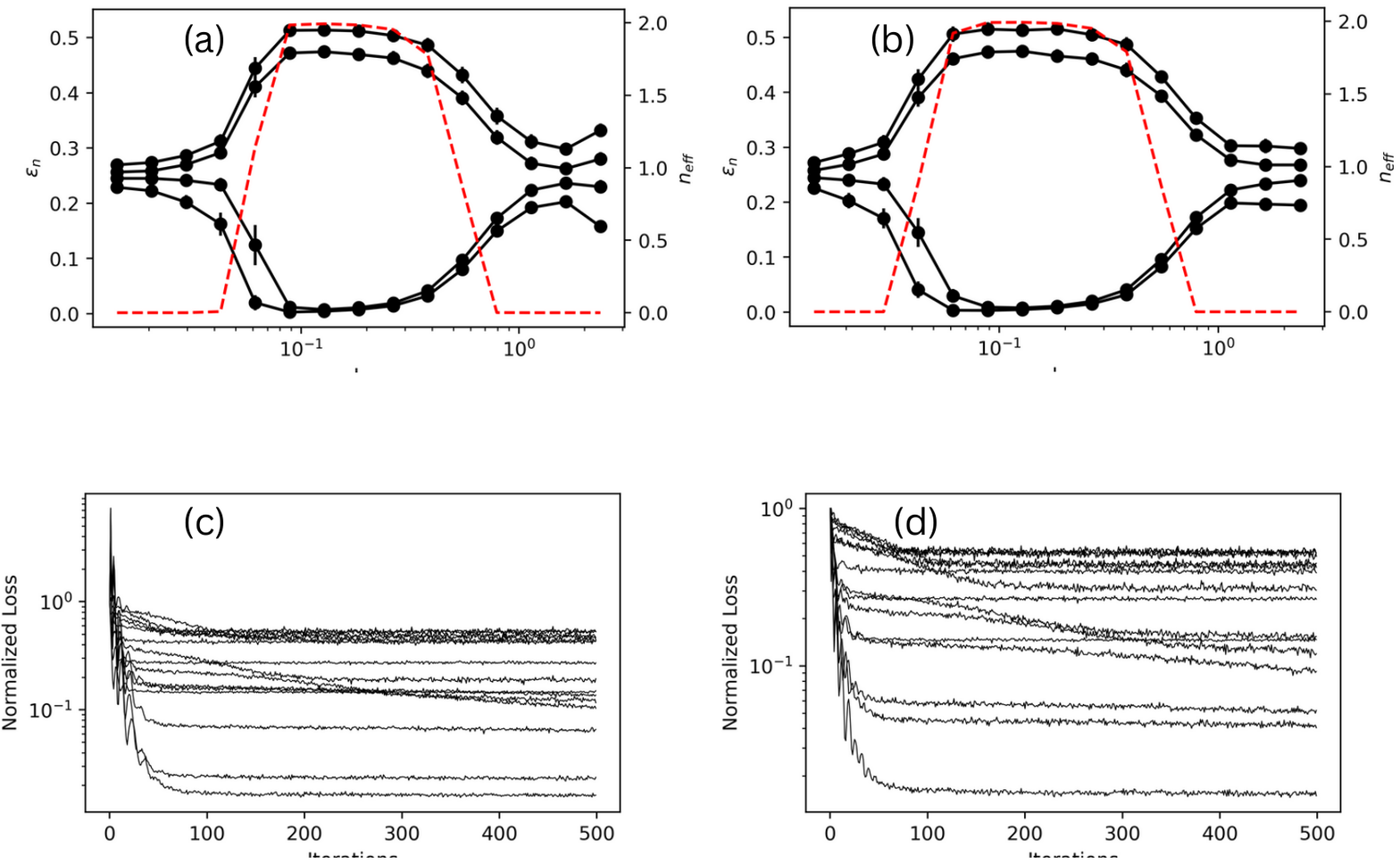}
\caption{ 
In the top graphs on panels (a) and (b), the Explained Ratio Diagrams depict the effective dimensionality of the manifold and the distribution of relative eigenvalues. Meanwhile, the bottom graphs on panels (c) and (d) illustrate the evolution of the loss function across epochs. These analyses are conducted on neural networks with distinct architectures: panels (a) and (c) correspond to a network with two layers, each containing 256 neurons, while panels (b) and (d) correspond to a network where the first layer has 128 neurons and the second layer has 512 neurons. These graphs provide valuable insights into the impact of varying neuron counts on both the characteristics of the learned manifold and the convergence behavior of the neural network during training. }\label{fig6}
\end{figure}

The initial architecture of the neural network comprised two layers, each consisting of 256 neurons. Referencing Fig.\ref{fig6} panel (a) and (c), which displays the corresponding Explained Ratio diagram and the associated loss function, it is evident that among the 15 neural networks implemented for various perturbation distances, only two exhibited loss functions below the threshold of 0.1. In response to this observation, architectural modifications were enacted.
The revised architecture maintains a two-layer configuration but with an adjustment in the number of neurons. Specifically, the first layer now comprises 512 neurons, while the second layer retains 256 neurons. As delineated in Fig.\ref{fig6}, the outcomes of this architectural modification reveal a notable decrease in the loss function of the optimal neural network. Notably, three neural networks now exhibit loss functions below the stipulated 0.1 threshold. Additionally, the Explained Ratio diagram demonstrates a more expansive and distinct shape, indicative of an improved performance compared to the previous configuration.

\section{Appendix}
The McMillan lens serves as a notable example of non-integrable optics, playing a crucial role in our study utilizing AI Poincaré analysis. In this experiment, an electron beam with a significantly larger radius than the original beam is injected into the accelerator ring. Following injection, Coulomb interactions between the injected and primary beams lead to the focalization of the latter.

\begin{figure}[t]
\centering\includegraphics[width=8cm]{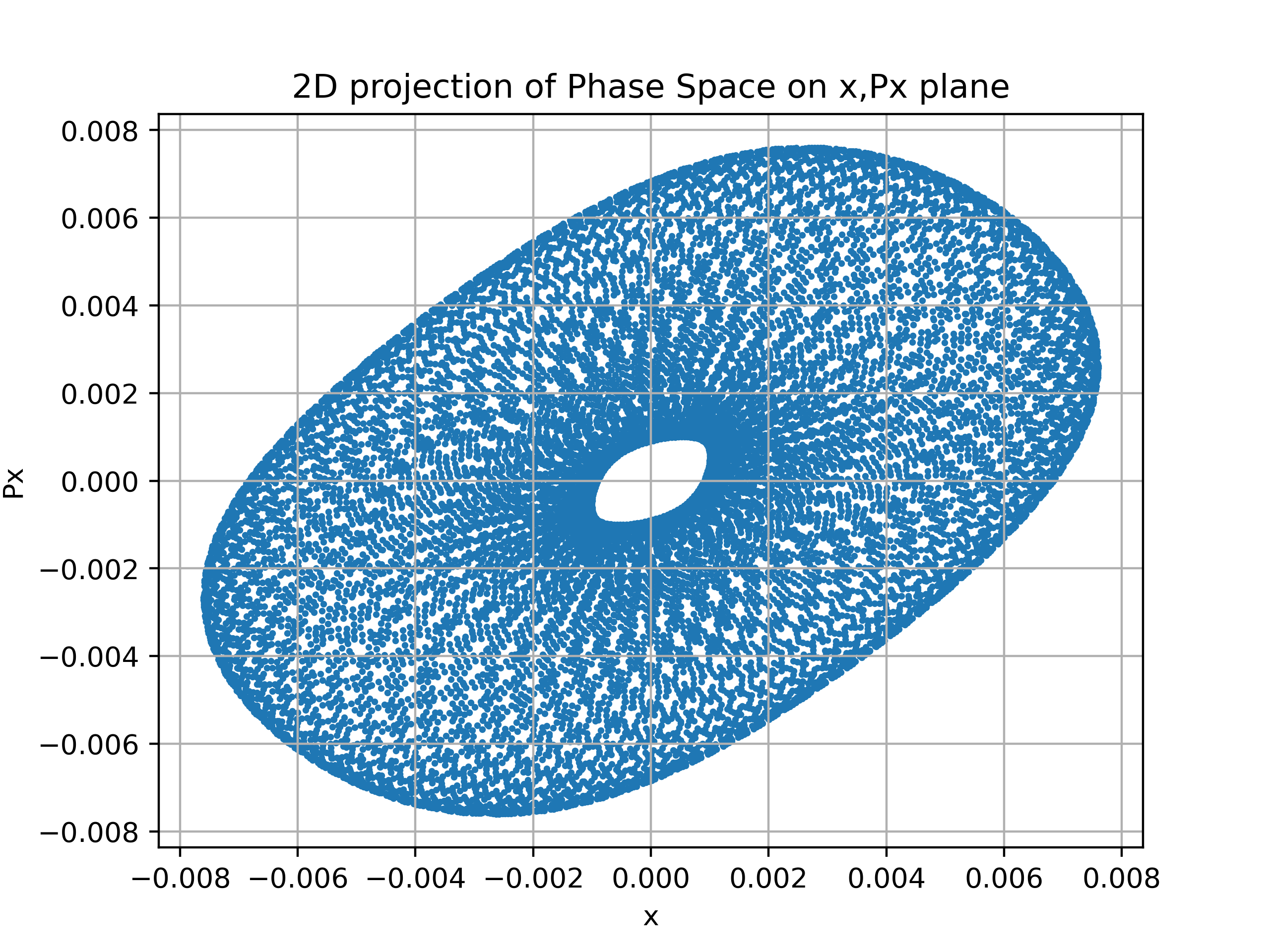}
\caption{ This figure presents a 2D projection of the McMillan map onto the phase space plane defined by the coordinates $x$ and $x'$. The initial conditions for this projection are specified as $x(0) = 0.001, x'(0) = 0, y(0) = 0.003,$ and $y'(0) = 0$. The visualization provides insight into the evolution of the system under these initial conditions, offering a concise representation of the McMillan map dynamics in the specified phase space. }\label{fig7}
\end{figure}

This system exhibits two degrees of freedom and precisely two conserved quantities, indicating integrability. Consequently, the phase space, which would typically be four-dimensional, effectively reduces to two dimensions due to the presence of conserved quantities. The system's dynamics are governed by the following transformation matrix \cite{lens}, which predicts the evolution of particle states turn by turn:

\[
\begin{bmatrix}
x \\ x' \\ y \\ y'
\end{bmatrix}_{a}
=
\begin{bmatrix}
0 & \beta & 0 & 0 \\
-\frac{1}{\beta} & 0 & 0 & 0 \\
0 & 0 & 0 & \beta \\
0 & 0 & -\frac{1}{\beta} & 0
\end{bmatrix}
\cdot
\begin{bmatrix}
x \\ x' \\ y \\ y'
\end{bmatrix}_{i}
\]

where $(x,x',y,y')$ denote the particle coordinates and momenta, the subscript i indicates the initial state and subscript a indicates the state after turn, and $ \beta $ represents the amplitude function. The resulting phase-space trajectories under the McMillan lens are illustrated in Fig. \ref{fig7}.

\section{Conclusions}
In summary, our research delved into understanding how AI Poincaré works, using it to test a system known as McMillan. We specifically looked at the limitations of AI Poincaré, examining the maximum perturbation distance where conserved quantities remain relatively constant across different neural networks. Our findings showed this range to be from 0 to 0.2, and even for larger perturbation distances, the points calculated by AI Poincaré stayed close to the initial points we chose. We also tested AI Poincaré on actual experimental data, confirming the conservation of at least one quantity.
Lastly, we tweaked the neural network's architecture, reducing loss functions and refining the shape of the Explained Ratio diagram. These changes improved the performance of AI Poincaré when applied to the McMillan system.

\bmhead{Acknowledgments}
We sincerely thank the Fermilab IOTA team for their invaluable support, expertise, and contributions to this research.

%\bmhead{Data Availability Statement}: The code and the datasets generated during and/or analyzed during the current study are available on GitHub.\footnote{\url{https://github.com/RocinantEL/ML-Application-chaos}}

\end{document}